# Efficient Creation of Behavior Models with Variable Modeling Depths Used in Digital Twins

Valentin Stegmaier [a,b,c,*], Walter Schaaf [c], Nasser Jazdi [b], Michael Weyrich [b]

[a] Graduate School of Excellence advanced Manufacturing Engineering (GSaME), University of Stuttgart, Nobelstraße 12, 70569 Stuttgart, Germany
[b] Institute of Industrial Automation and Software Engineering (IAS), University of Stuttgart, Pfaffenwaldring 47, 70569 Stuttgart, Germany
[c] J. Schmalz GmbH, Johannes-Schmalz-Straße 1, 72293 Glatten, Germany

*Corresponding author: Tel.: (+49)7443-24037443, Email: valentin.stegmaier@gsame.uni-stuttgart.de

*Abstract* — **Behavior models form an integral component of Digital Twins. The specific characteristics of these models may vary depending on the use case. One of these key characteristics is the modeling depth. Behavior models with a lower modeling depth depict the behavior of the asset in an abstract way, while those with a higher modeling depth depict the behavior in detail. Even if very detailed behavior models are flexible and realistic, they also require a lot of resources such as computing power, simulation time and memory requirements. In some applications, however, only limited resources are available. The automated creation of Digital Twins is of crucial importance for their widespread use. Although there are methods for the automated creation of behavior models for Digital Twins with a specific modeling depth, there is currently no method for the automated creation of behavior models with varying modeling depths. This article presents such an approach and demonstrates its advantages using two industrial use cases. It is demonstrated that the automatically created behavior models of lower modeling depth yield results that are almost identical to those of models with a higher modeling depth, but with significantly reduced computing time and required memory. This enables the efficient use of behavior models in a variety of use cases, regardless of the availability of resources.**

## I. Introduction

The increasing digitalization and automation of discrete production systems represents a crucial strategy for sustaining the economic viability of industrial facilities situated in regions characterized by relatively high wage costs [1]. The Digital Twin (DT) represents a pivotal element in the transition towards digital and highly automated production systems. [2–4]. A consistently available digital representation of physical assets can save costs and time in design, development and commissioning as well as in the operation of production facilities [5, 6]. Literature presents a variety of different concepts for the Digital Twin [7, 8]. In this paper, we apply the definition of Ashtari et al. [7] which defines a Digital Twin as a virtual representation of an object, frequently referred to as an asset, that allows to represent the assets static and dynamic behavior [9]. The Digital Twin contains all relevant models of the asset, along with all relevant data from the various stages of its lifecycle. It allows for the simulation of the physical behavior of the object within a virtual environment, and is always synchronized with the asset [9]. In addition to further peripheral properties, such as a unique ID or a Digital Twin version management system, data and models represent two of the core aspects of a Digital Twin [7, 10]. Among these models, behavior models represent an important group of models [11, 12]. They provide a description of the behavior of components or systems consisting of multiple components and are typically executable [13]. Colloquially speaking, they are also called simulation models [13]. Behavior models are of significant importance in a multitude of Digital Twin applications, including virtual product design, virtual product validation and virtual commissioning (VC) and are frequently used in the contemporary development process [11, 14, 15]. In addition to a significant variety in the specifications for the scope of the behavior models and the disciplines under consideration, there are also notable variations in the required modeling depth (MDT) depending on the use cases [10, 16].

The MDT is defined as the level of abstraction associated with the representation of a model, which can range from an abstract or idealized representation to a highly specific and precise representation [17]. There are various approaches to classify the depth of modeling. In this article, a 5-level approach, as illustrated in Figure 1, will be used [17]. To illustrate the difference between the MDT levels, some exemplary industrial use cases are shown in Figure 1 too.

| Degree of abstraction / Modeling depth | | | Exemplary industrial use case |
|---|---|---|---|
| Time and space discrete | MDT 1 | Discrete behavior | IO-Tests |
| Time continuous and space discrete | MDT 2 | Discrete temporal behavior | Virtual commissioning |
| | MDT 3 | Continuous simplified physical behavior | Condition Monitoring |
| | MDT 4 | Physical non-spatial behavior | System layout/ configuration |
| Time and space continuous | MDT 5 | Physical spatial behavior | Component optimization |

Figure 1: MDTs and exemplary industrial use cases [17]

The 2 highest MDTs only describe discrete behavior, without intermediate states. The difference between the MDT 1 and 2 is the inclusion of time delays in the latter. In contrast to MDT 1 and 2, 3 describes continuous behavior. However, it is modeled in a highly simplified manner and is only capable of coping with very simple equations. This changes with MDT 4, which models the physical behavior with more sophisticated equations, e.g. using differential equations. The only simplification of this MDT is the neglect of local expansion, which is considered in MDT 5. Some of the behavior models are limited to discipline-specific tasks, such as optimizing the fluidic efficiency of pneumatic vacuum generators using FEM simulations (finite element method)[18]. The MDT of such models may be too substantial to enable simulations of entire systems with multiple components within the time constraints typically associated with numerous Digital Twin applications [17, 19]. Behavior models with a MDT of 4 or lower are frequently used for simulations of systems with multiple components and represent the primary focus of this article [17].

A widespread use of Digital Twins requires their creation with minimal effort, regardless of the MDT. While a large part of the relevant data and models of a Digital Twin are already available in companies in most cases, behavior models must often still be created in the process-relevant MDT [20]. One approach to create such behavior models with minimal manual effort is shown in Figure 2.

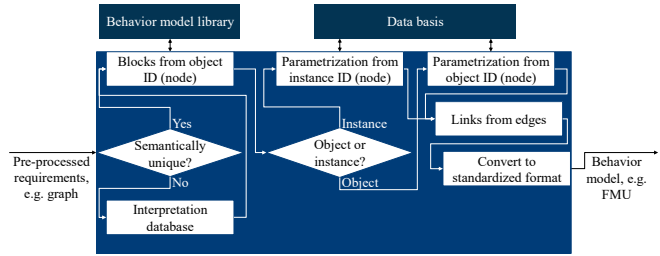

Figure 2: Method for automated creation of behavior models with MDT 4 (AutoBMC method) [20]

As inputs, information regarding the structure of the behavior model can be represented in the form of a graph, which is utilized to extract the building blocks from the behavior model library. Following the parameterization of the extracted behavior models, the building blocks are linked using information from the graph. The behavior models can then be converted into a desired standardized format such as Functional Mock-Up Units (FMUs) or exported directly.

With this approach, the creation of behavior models is automated, but only for a defined modeling depth. The goal of this paper is to present a method for the automated creation of behavior models with different modeling depths as shown in Figure 3. This enables the efficient use of behavior models in a variety of use cases, regardless of the availability of resources such as such as computing power, simulation time and memory requirements.

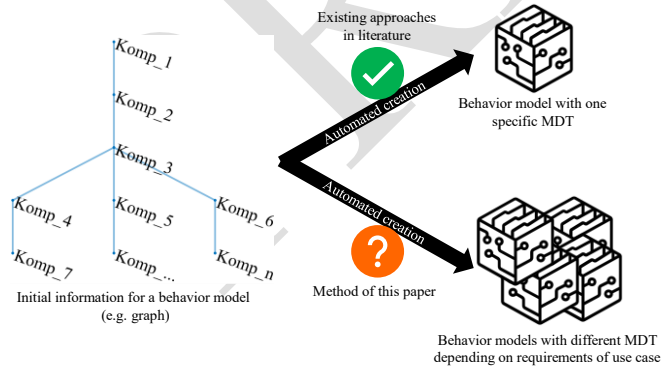

Figure 3: Goal of this paper

The following section provides an overview of existing approaches from literature. Subsequently, section 3 outlines the method to create abstracted behavior models. This is implemented in section 4 and evaluated based on two industrial use cases in section 5. The article concludes with a discussion with a summary and an outlook.

## II. STATE OF THE ART

In the following, the required fundamentals for discrete production systems are first discussed before relevant approaches for abstracting the modeling depth from the literature are listed.

Typical discrete manufacturing processes can be classified into a discrete number of phases. Vacuum gripping systems serve as an example for this [21]. The process of such an automation component can be divided into the phases as shown in Figure 4 [21, 22].

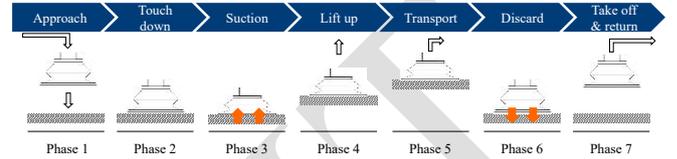

Figure 4: Phases of a typical handling process of a vacuum gripping system based on [21]

The approach (phase 1) and touchdown (phase 2) of the vacuum gripping system on the object to be handled marks the beginning of the process. The object is then sucked (phase 3), lifted (phase 4), and transported (phase 5). Once the object to be handled has reached its target position, it can be discarded (phase 6), and the vacuum gripping system can be removed and returned to its starting position (phase 7).

The abstraction of behavior models is a topic that has been explored from a variety of perspectives in literature. These approaches can be roughly classified into two categories: Model-based and data-driven. [23–25]

Model-based approaches are based on a mathematical or physical understanding of the models. In this domain, there are highly specialized methods tailored to specific categories of models. One illustrative example is the Craig-Bampton method, which reduces the number of degrees of freedom in finite element models to create an abstracted model [26]. In the context of linear systems, techniques such as linearization, linear parameter variation or balanced truncation can be utilized [25, 27–29].

The group of data-driven approaches utilize the input and output data of the original detailed behavior model. Potential methods for the static abstraction of the MDT include interpolation and the utilization of lookup tables. For dynamic, abstracted behavior models, Long Short-Term Memory (LSTM) networks, forward neural networks, or neural networks with ordinary differential equations can be used. [25, 30–32]

Examples of implemented approaches for the abstraction of behavior models include "RBmatlab" [33], "model reduction inside ANSYS" [34], "pyMOR" [35] and "MORLAB" [36]. These are frequently functions, extensions or libraries for specific simulation programs. During the process of abstraction, decisions must be made regarding the subsequent use of the abstracted behavior models. Additionally, there is not yet a unified method for the abstraction of all potential behavior models. [23]

In the process of behavior model creation for component manufacturers, it is not always feasible to assume that the mathematical or physical understanding can be accurately captured, given the high diversity observed in the behavior models used. Additionally, these models are not always linear systems, which further limits the applicability of purely model-based approaches as a means of abstraction. However, data-driven techniques are promising for the abstraction of behavior models. Nevertheless, no approach could be found in literature that uses the subdivision of discrete manufacturing processes for abstraction.

***Deduced requirements for the concept:*** Based on the first 2 sections of this article, the following conclusions can be drawn for the concept:

- Behavior models as a central component of Digital Twins are required for different use cases in different MDTs.
- In literature, there are approaches for the automated creation of behavior models in one special MDT.
- Behavior models of MDT 1, 2, 3 and 4 are particularly relevant for use cases that consider entire components and systems.
- The workflow of typical discrete manufacturing processes can be described by repeating cycles, which can be divided into a few discrete phases.
- Existing approaches from literature are very generic, which means that they can be used in a variety of ways on the one hand but are highly complex on the other hand.
- Data-driven approaches are a promising way of abstracting behavior models however no approach could be found in literature that performs the abstraction of the MDT using discrete states.

Based on these deductions, a concept for the efficient creation of behavior models at 4 different MDTs will be presented below.

### III. CONCEPT

This chapter first compares different options for creating behavior models at different modeling depths before presenting a method based on the preferred option. In principle, behavior models of MDT 1, 2, 3 or 4 can be created manually from behavior models of higher MDTs or from scratch. However, this requires a simulation expert with a high level of understanding of the system to be modeled, is time-consuming, and error-prone. A common mistake in abstracted behavior modeling is neglecting or forgetting special and rarely occurring effects. For this reason, only automated methods will be considered below.

#### A. Options to create behavior models with different MDTs

The automatically created behavior models with a specific MDT using the approach from Figure 2, a behavior model library in this particular MDT is needed. However, since the library only contains behavior models of one MDT, the method in Figure 2 can only be used to create behavior models of one MDT. If behavior models of different MDTs are to be created, there are 2 main approaches that can be taken:

- *Option 1:* The approach described in Figure 2 is extended to include behavior model libraries corresponding to the specified MDTs. In this case, a total of 4 behavior model libraries would be required, covering MDT 1, 2, 3 and 4.
- *Option 2:* Offering a single behavior model library. However, the behavior models created with this library are abstracted towards the desired MDT with an additional method once they have been fully created.

The 2 options have specific advantages and disadvantages. The preliminary work involved in creating the behavior model library is considerably more extensive for option 1. However, option 1 does not require the development of an additional method, and the actual creation of the behavior models is more straightforward with less effort. Option 2, on the other hand, requires less initial effort to create the libraries, but requires an additional method for abstraction. However, the creation of behavior models for systems with an MDT of 1, 2 or 3 is problematic with option 1, as significant discrepancies between the modeled and actual behaviors can arise in such cases. The issue is demonstrated using a small-scale experimental setup. This setup comprises a pneumatic vacuum generator, a reservoir, and a pressure sensor, shown as test setup 1 on the left in Figure 5. The setup is replicated for test setup 2, except for the reservoir, which is replaced by a hose, shown on the right in Figure 5. The reservoir and hose represent the internal volume of a vacuum gripping system in simplified form.

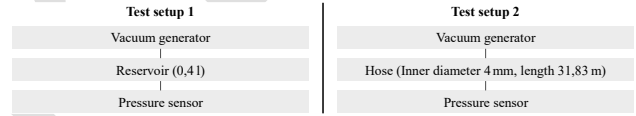

Figure 5: 2 test setups for an identical system volume

The reservoir has a volume of 0.4 l, and the hose (with an inner diameter of 4 mm) has a length of 31.83 m. This indicates that the internal volumes of the hose and reservoir are identical. Equation (1.1) may be used to determine the evacuation time ($t$) as a function of the internal volume ($V$) of a system, the maximum suction capacity ($S$) and the maximum vacuum ($p_v$) of the vacuum generator and the target vacuum ($p_0$) [37]. This approach is subject to abstractions and enables the mapping of temporal behavior but does not calculate intermediate states. For these reasons, equation (1.1) can be classified as MDT 2.

$$t = \frac{V}{S_N} \cdot \ln\left(\frac{p_0}{p_v}\right) \qquad (1.1)$$

The evacuation time to a target vacuum is a crucial performance indicator in vacuum gripping systems, as it affects the overall cycle time [21]. The exemplary formula for a behavior model of MDT 2 yields a consistent evacuation time for both test setups depicted in Figure 5. However, in contrast to MDT 2, the discrepancy between the 2 evacuation times in reality is considerable. For an evacuation process up to 700 mbar,rel the time taken for the reservoir with the selected vacuum generator (SBP 20 [38]) is 0.36 s, whereas the time taken for the hose is 2.32 s, representing a discrepancy of a factor of 6.4. This discrepancy is primarily caused by the flow resistance, which is not taken into account in the abstracted equation (1.1). Besides the flow resistance, numerous other effects are not considered in the MDT 1, 2 and 3. For this reason, MDT 4 is recommended for the development of sufficiently precise behavior models of entire systems. If behavior models with a lower MDT are required, these should be abstracted from the behavior model of the overall system.

Accordingly, option 2, comprising a behavior model library and a method for abstracting other MDTs, has been selected for the automated creation of behavior models with variable MDTs. Considering the simplifications of MDT 1, 2 or 3 and their inherent limitations in terms of system configuration from individual components, the fourth MDT is selected for the behavior model library. Behavior models of MDT 4 are created and parameterized using this library. These may then be abstracted to MDT 1, 2 or 3. This process is shown in Figure 6 as the overall concept of this article.

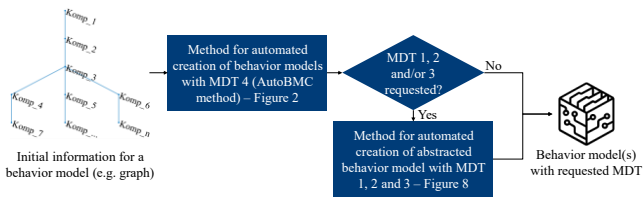

Figure 6: Overall concept of this article

In contrast to the abstraction of the MDT, it is generally not possible to automatically derive behavior models of MDT 5 from behavior models of MDT 4 without further information.

*B. Method to abstract the MDT*

The method for the automated abstraction of the MDT is based on the segmentation of typical automated processes of discrete manufacturing into states and transitions. Figure 4 provides an illustrative example of this concept. To progress from one state to the next, specific conditions must be satisfied. If a condition is met, a transition can execute. It is precisely this structure of states and transitions with conditions that can be well modeled using state machines. However, it should be noted that an abstracted behavior model does not always have to represent all process steps of a component. There are also instances where a component is inactive, and the considered outputs remain constant. An example of this are phases 7, 1, and 2 in Figure 4, where there is no significant change in the vacuum curve of the vacuum gripping system.

To create an abstracted behavior model, it is at first necessary to capture the relevant information for the state machine. The behavior model of MDT 4 is used as a black box for the purpose of detecting the states and transitions. Only the inputs of the model are applied, and the outputs are analyzed with the resulting values. This information can then be used to create the abstracted models of MDT 1, 2 or 3. For MDT 1, the states of the system are created as states of the state machine and the conditions between the states are added according to the behavior of the component or system. This structure is illustrated on the left-hand side of Figure 7. However, this structure does not permit the mapping of temporal behavior, as the state changes as soon as a condition for a transition is met.

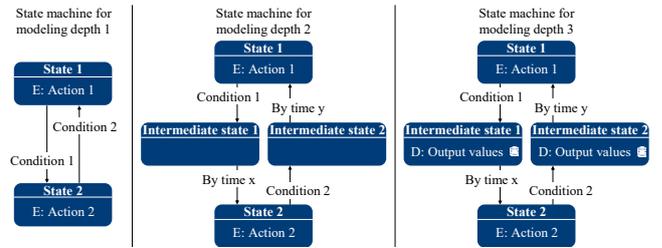

Figure 7: Structure of the state machines for the different MDTs

To incorporate a temporal element, it is necessary to construct behavior models of MDT 2. These are fundamentally based on the state machines of MDT 1 but supplement the transitions between 2 states with at least 1 additional state, which is referred to as an intermediate state. The intermediate state is initiated directly upon the fulfillment of the transition condition. In this intermediate state, however, there are no actions that affect the outputs of the behavior model. Such actions are only carried out in the actual target state. The target state is initiated after a defined period has elapsed. This is achieved through a transition between the intermediate state and the target state, with the defined time acting as a condition for the final state to become active. This structure enables the input signals to exert temporal effects on the output signals, provided that specific conditions are met. The structure is illustrated in the middle of Figure 7.

Nevertheless, this structure does not permit the mapping of continuous processes. The structures for MDT 2 permit the creation of discrete end states. However, they do not support the generation of continuous intermediate states. To facilitate the output of signals not only in a discrete manner, but also with the inclusion of continuous intermediate values, the state machine of MDT 3 has been developed. This is based on the state machine of MDT 2 but extends it to include an action in the intermediate states. In the event of their activation, values are transmitted in the intermediate states according to a defined cycle. It is essential to determine the cycle in advance. Each intermediate state comprises an array of values, the size of which is determined by dividing the transition time from the intermediate state by the cycle time. Upon completion of the transition time, the intermediate state is exited, and the output is set to the constant value of the active state. This structure allows for the definition of continuous curves for output signals, dependent on the input signals.

To create state machines of MDT 1, 2 or 3, it is necessary to determine the states, transitions, conditions for the transitions, time delay of the transitions, and values for the transition times. The process to identify the relevant information is shown in Figure 8.

Before the information can be identified, relevant inputs and outputs must be defined, and the interesting value ranges of the inputs and the relevant steps must be specified. In principle, the method described below can be used in the absence of information regarding the value range and the relevant steps. Nevertheless, this would lead to a considerable increase in the number of iterations, as all potential values for the input would then have to be evaluated with the smallest relevant step size. To illustrate, consider a typical input signal in automation technology, which has two states: high level and low level. In the case of these two levels, for instance, the values 0 V to ground and 24 V to ground are used. If the method is provided with information regarding the value range but not the step size, it would be necessary to evaluate the entire value range, for example, in increments of 1 V. This would necessitate the completion of a total of 25 iterations, in comparison to just 2 iterations if the step size is known.

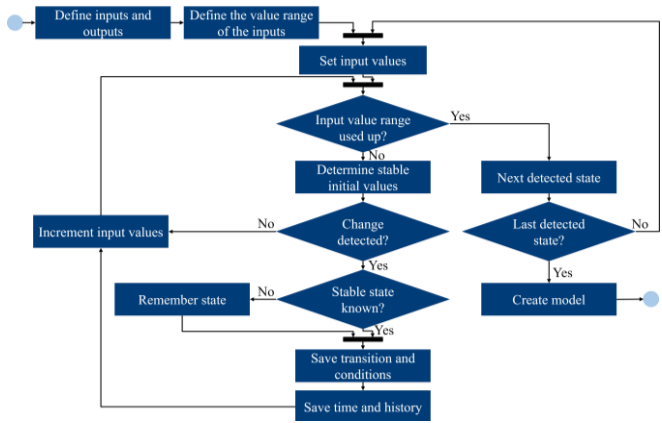

Figure 8: Method for automated creation of abstracted behavior model with MDT 1, 2 and 3

By defining the value range for an input, the input can be classified as either relevant or irrelevant for the abstracted behavior model. If a single value is assigned to an input, this value is assumed to be constant within the behavior model, and the input is subsequently excluded from the abstracted behavior model. This procedure can be used to specify the number of relevant inputs within the abstracted behavior models. The inclusion of additional relevant inputs, accompanied by correspondingly more relevant values per input, results in an increased execution time for the method.

Once the value range and increments have been defined, an initial simulation is conducted with the first value for each input. The simulation time for a set of values for the inputs is also a parameter that influences the performance of the method. Selecting a longer simulation time for each iteration allows for the more reliable detection of a stable state, as it allows for the consideration of longer transient processes in the system. However, this also results in a longer execution time for the method.

Once the initial simulation is complete, the stable values of the outputs for the initial state are determined. These are the values that are set after a specified time and then remain constant thereafter. From this point onward, a series of input values may be attached to those initial input values. By adding new input values to the initial input values, the simulation always starts with the initial state, to which the new input values are then applied. Following the completion of the simulation, a comparison is made between the current outputs and the output values of the initial state. If no change in output values is detected, the subsequent set of input values can be established, thereby initiating a further simulation. In the event of a change being identified, it is first necessary to determine whether this state has already been observed. If not, the state is stored as a new state in the state memory. Regardless of whether the final state reached by the simulation is already known, the transition itself, the conditions that must be met for it to occur and the time of all output signals associated with this transition are stored in the transition memory. The next set of input values is then set, and the next simulation is initiated. This process is repeated until all previously defined input values have been evaluated. If so, the next state is defined as the initial state, and all sets of input values are iterated through again for this state.

The process for detecting all states and transitions is complete when all input values have been iterated through for all found states as start states. At this point, it can be ensured that all relevant information for the creation of the abstract behavior model has been collected. The corresponding model can then be built according to the structures described in Figure 7.

The process to create the behavior models at the specified MDT is also automated. The initial step is to create all recognized states with their corresponding actions. The action then sets the output signal to the previously recognized value. If intermediate states are required in the desired MDT, these are also added. Subsequently, all transitions from the transition memory are added with the corresponding conditions. The transitions from the states are provided with the corresponding recognized conditions, while the transitions from the intermediate states are provided with the recognized time delays. This guarantees that the intermediate state will remain active for the duration specified in the intermediate state memory. Ultimately, if necessary, the progressions for the output of the signals in the intermediate states of MDT 3 are stored in variables within the model. The behavior model created in this manner can be encapsulated, thereby offering the defined inputs and outputs from the corresponding behavior model of MDT 4.

## IV. REALIZATION

MATLAB with various extensions is used for the implementation. It facilitates the straightforward integration of simulations with logic programming and a user interface. The AutoBMC implementation from [20] is used for the automated creation of the behavior models in MDT 4. The automated creation of a behavior model from corresponding input information is illustrated in Figure 9. A vacuum gripping system, used in a multitude of automated processes for the handling of diverse components, is taken as a representative industrial use case scenario [39].

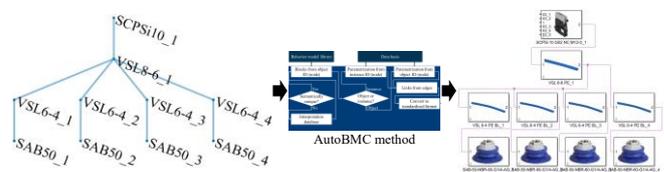

Figure 9: Automated creation of an exemplary behavior model with MDT 4 from a graph using the AutoBMC method [20]

To abstract this behavior model, it is first necessary to define the relevant inputs and their value ranges and step sizes. In this paper, a user query is used to identify the relevant inputs and their associated value range. In this user query, the possible values for each input can be entered, separated by commas. If an input is to be excluded from the model or maintained at a constant value, the user is required to enter the desired constant value for that input. In this manner, the system is able to identify that this input should remain unchanged throughout the iterations and thus excludes this input from the resulting abstracted behavior model. Furthermore, the time to be used for the input signals until a stable state is reached must be specified. It is preferable to select a longer time than a shorter one, as a shorter time may not allow the system to reach a stable state. In this work, a time span of 3 s is assumed for the behavior models since this value has been empirically proven to be efficient. This means that an input signal is present at the input of a behavior model for 3 s before switching to the next set of input values. ve.

The subsequent stage is to identify the initial state of the outputs of the behavior model with MDT 4. This refers to the state that occurs when value of 0 is applied to all inputs, which serves as the primary output state. Subsequent states are then compared with this initial state to detect any changes. An illustrative example of such a progression of input signals and the corresponding output signals returned by the behavior model is presented in Figure 10.

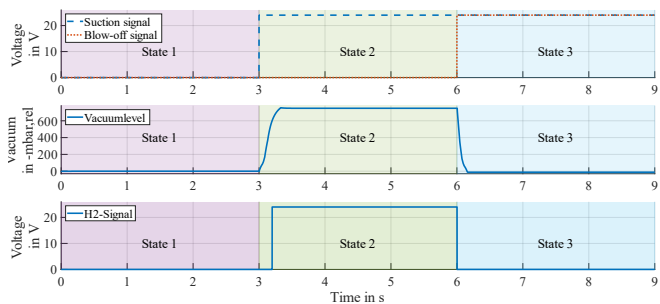

Figure 10: Example of input signals (top) and output signals (middle and bottom) for detecting states and transitions

It illustrates the 2 input and 2 output signals for the behavior model depicted in Figure 9 over a total of 9 s. As previously stated, one set of input signals is applied for a period of 3 s. If both input signals are equal to 0, this results in the initial state. The combination of outputs is defined as state 1 and is depicted in purple; it represents the initial state. In this state, both the signal for the vacuum value and the signal for the switching threshold (H2) output a value of 0. However, if an input signal with a value of 24 V (e.g. suction active) is applied, the model responds and a new state with the number 2 is created. This is illustrated in green in Figure 10. In this state, a vacuum value of approximately 750 mbar,rel and a voltage signal of 24 V for H2 are applied to the outputs. If, from state 2, a value of 24 V is applied for the suction and blow-off signal, a further state with the number 3 results. Although this is similar to state 1 in terms of the signals, it provides a value of -12 mbar,rel for the vacuum output signal.

In accordance with this method, all potential combinations of input signals are applied to all identified states. This process enables the reliable detection of all states and transitions. Multiple transitions with different conditions can exist between two states. Both the states and the transitions are stored with all pertinent information in an array. The arrays created for the detected states and transitions are presented in Table 1 and Table 2 for the behavior model in Figure 9.

Table 1: Values for states of the behavior model from Figure 54

| Name state | Number | Stable output values | Input values to reach state |
|---|---|---|---|
| 1 | 1 | 0, 0 | 0,0 |
| 2 | 2 | 749, 24 | 0, 0; 24, 0 |
| 3 | 3 | -12, 0 | 0, 0; 0, 24 |

For the 3 stable states detected, the values of the outputs detected as stable by the method are listed along with their name and state number. These are compared with possible new states to determine whether the state is new or already known. The values required for the inputs of each state to reach the state are also given. There are some states for which a particular sequence of input values must be applied to the behavior model to reach the state. This is visualized by multi-line input values. In addition to the states, the transitions are shown in Table 2.

Table 2: Values for transitions of the behavior model from Figure 54

| Start state | Values input | Target state | Time for each output in ms | Output values during transition |
|---|---|---|---|---|
| 0 | 0,0 | 1 | - | - |
| 1 | 24,0 | 2 | 319, 199 | [1x2] |
| 1 | 0,24 | 3 | 103, 0 | [1x2] |
| 1 | 24,24 | 3 | 103, 0 | [1x2] |
| 2 | 0,24 | 3 | 161, 0 | [1x2] |
| 2 | 24,24 | 3 | 161, 0 | [1x2] |
| 3 | 0,0 | 1 | 0,0 | [1x2] |
| 3 | 24,0 | 2 | 319, 198 | [1x2] |

For these, the start and target states for the respective transition are given, as well as the values for the inputs that must be applied to trigger the respective transition from the corresponding state. In addition, two further columns of information are listed for the abstracted behavior models of MDT 2 and 3. Firstly, the time for each output signal to change state after an input signal has been applied. Secondly, the exact course of the output values during this transition time with the specified time step width. The output values during the transition are stored in a cell array in MATLAB, as this makes it easy to combine vectors of different lengths. The next step is to create the state machines using the information from the states and transitions. The user interface can be used to specify which MDT (MDT 1, 2 or 3) is to be created. For all MDTs, the first step is to create an empty model for a state machine in MATLAB. The states from the state array are iteratively inserted in the next step. The states are numbered consecutively. In the state itself, the corresponding outputs are set as previously detected. In the next step, the detected transitions between the created states are added accordingly and these are supplemented with the detected conditions. Additional intermediate states with corresponding outputs can be added if required. The result is a model with inputs and outputs. In this way, the behavior models of modeling levels 1, 2 or 3 can be created according to the specification. The results of this creation using the example of the behavior model from Figure 9 are shown in Figure 11.

In the upper part the 3 states from Table 1 for MDT 1 are shown. There are a total of 8 transitions between them. These 3 states are supplemented by intermediate states for MDT 2. At MDT 3 (Figure 11 bottom), the intermediate states are supplemented by the continuous output of values shown in column 5 on the right in Table 1.

The 3 states shown in Table 1, Figure 10 and Figure 11 correspond to the inactive phase of the gripping system (phases 1,2 and 7), the phase with active vacuum (phases 3,4 and 5) and the discard phase (phase 6) of Figure 4. This confirms as stated above an abstracted behavior model does not always have to represent all states of a component or system.

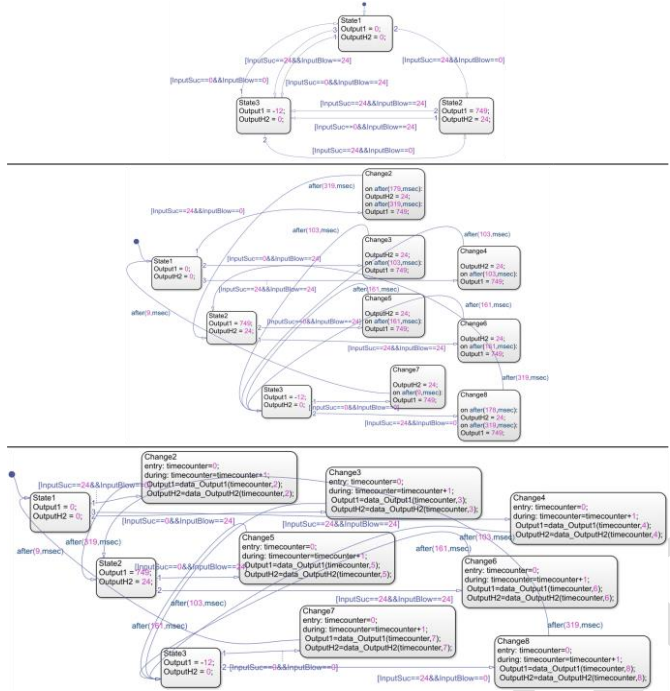

Figure 11: Behavior models of MDT 1 (top), 2 (middle) and 3 (bottom) realized in MATLAB of the behavior model shown in Figure 9

The behavior models created in this way for MDT 1, 2 or 3 can be saved and used directly as state machines. This makes it easy to view the logic stored in them, but requires the same development environment, in this case MATLAB with the appropriate extensions. The state machines can also be exported as FMUs. This allows for tool-independent use of the created behavior models and makes it easier to hide the logic stored in the behavior model. This allows companies to share their behavior models across company boundaries without directly exposing explicit knowledge that is worth being protected.

## V. EVALUATION

The evaluation is carried out using 2 industrial use cases. After describing them, the results achieved are presented and the section is concluded with a discussion. Vacuum gripping systems are used for the evaluation since such systems are widely used in the industry due to their robustness and ease of implementation compared to other gripping technologies [40–42]. The discipline considered in this article is fluidics, as this is central to the relevant behavior of vacuum gripping systems [20]. The behavior models are evaluated in 2 stages. In the initial phase, the behavior models created with the AutoBMC method are compared with measurement data to confirm their alignment with the actual behavior of the systems. In the second step, the abstracted behavior models can then be compared with those of MDT 4. If the behavior models of MDT 1, 2 or 3 are found to align with those of MDT 4, it is assumed that the abstracted behavior models will reflect the actual behavior of the systems.

### A. Use case 1: Vacuum gripping system for automotive body shop

The first use case is illustrated by considering the example of a vacuum gripping system in the context of an automotive body shop. Vacuum gripping systems are used, among other things, for stacking metal sheets, loading and unloading forming presses and transferring between forming presses. The used system is shown in Figure 12. This represents a simplified system of a typical vacuum gripping system but has all the relevant aspects and components of a typical vacuum gripping system such as a vacuum generator, 5 hoses, a distributor and 4 vacuum suction cups.

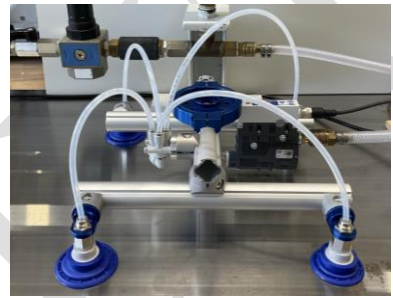

Figure 12: Vacuum gripping system of use case 1

A graph containing the necessary information is available for the creation of the system. The behavior models are then created by the AutoBMC method in MDT 4 using graphs. To evaluate the behavior models of MDT 4, it is essential to conduct a comparison between the models and the actual system. To this end, the vacuum gripping system depicted in Figure 12 is to be measured. The vacuum value at the vacuum generator, which is recorded with an external vacuum sensor and a programmable logic controller (PLC) type Beckhoff CX-9020, is utilized for this purpose. The measured values are recorded at a sampling rate of 1 ms. The measurement cycle lasts a total of 11 s. The vacuum gripping system is in the suction state for 5 s, in the release state for 1 s, and in the passive state for 5 s. The cycle is repeated a total of 30 times for the measurement, allowing any statistically distributed deviations in the measurement setup to be detected and quantified. The relevant part of a measurement cycle obtained from the test setup is presented in blue in Figure 13 in an overview (left) and zoomed in (right).

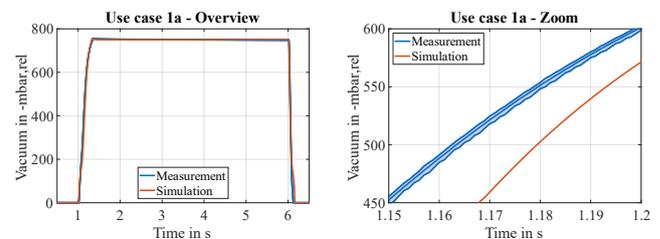

Figure 13: Comparison of measurement and simulation for use case 1a in an overview (left) and zoomed in (right)

The figure illustrates the maximum, mean, and minimum values of the vacuum curve. It is challenging to differentiate between the average, maximum, and minimum curves in the overview plot. Even in the zoomed in graph only a slight deviation can be noted. This suggests that the scattering is minimal and that the process is stable. The measurements are then compared with the results of the simulation. These are illustrated in orange in Figure 13. The curves of the measurement and the simulation are almost as congruent as the curves for minimum, maximum and mean value of the measurements. The slightly larger deviation only becomes clear with the zoomed in plot on the right in Figure 13. Here one can see a certain deviation between measurement and simulation. A maximum deviation of 80 mbar,rel can be determined between measurement and simulation in the area of the rising vacuum. During the constant vacuum phase, the deviation is less than 12 mbar,rel. A slightly larger deviation is evident in the blow-off area, with a maximum of 122 mbar,rel occurring shortly before the vacuum drops completely. Nevertheless, a highly satisfactory correlation between the measurement and simulation results can be stated for use case 1.

The behavior model of the vacuum gripping system, created with the AutoBMC method and subsequently compared with the measurement, has been assigned to MDT 4. For the behavior models of MDT 1, 2 and 3, correspondingly large simplifications are made depending on the MDT. To determine the correspondence between the detailed and simplified behavior models for use case 1, Figure 14 is presented.

The upper graph of Figure 14 illustrates the course of the input signals that are applied to the behavior model. The signals in question are of a binary nature and exhibit a maximum voltage level of 24 V. It can be observed that upon the application of a positive suction signal, the vacuum rises to an approximate level of 750 mbar,rel. The output curve is dependent on the MDT. It is created without a time delay (MDT 1), with a time delay but without a continuous curve (MDT 2), or with a time delay and a continuous curve (MDT 3). This demonstrates that the signals of MDT 3 and 4 have a high degree of overlap. When only the 2 input signals are utilized as variables, MDT 3 show the same behavior as MDT 4. This phenomenon also occurs regarding the range of the falling signal. In this case, the behavior models with a lower MDT reproduce the behavior of MDT 4 in accordance with their inherent limitations.

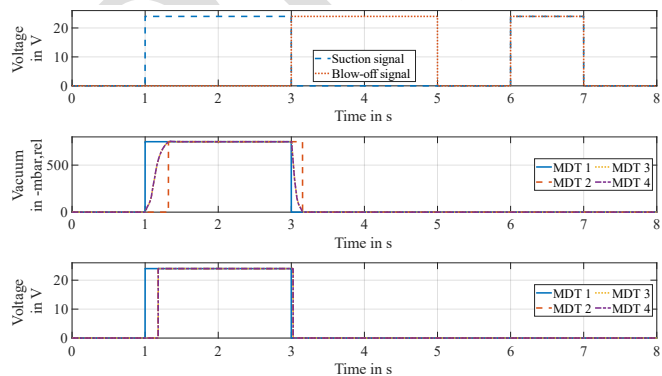

Figure 14: Behavior of the behavior models of MDT 1, 2 and 3 in comparison with MDT 4 for the signal H2 (bottom) and the system vacuum (middle) depending on the input signals suction and release (top)

This is also evident when examining the digital output signal H2 in the lower graph of Figure 14. Here, there is also a high degree of correlation between the abstracted behavior models of MDT 1, 2 and 3 and the behavior model of MDT 4, given the simplifications inherent to each. However, the lack of intermediate steps between the signal's discrete points hinders the ability to differentiate between the behavior models of MDT 2 and 3.

The capacity of simplified behavior models to represent not only a typical evacuation cycle but also the underlying logic of such a system is evident from the sixth second of Figure 14. When a positive signal for suction and blow-off is applied simultaneously, the blow-off signal assumes a dominant role. This indicates that the evacuation process does not take place, and the vacuum level remains constant. Consequently, there are no changes in the vacuum curve (middle) or in the signal H2 (bottom). This behavior is also reflected in the behavior models of MDT 1, 2 and 3, which demonstrates that they adopt the behavior of MDT 4 in accordance with their limitations, resulting in slight deviations from real life behavior.

The significant benefits of behavior models with a reduced MDT are evident when the time required for pure execution and for execution including compilation are considered. For this purpose, the models are executed 30 times (with compilation) and in MATLAB's "Fast Restart" mode (without compilation), whereby the required times are determined using internal MATLAB functions. To minimize external influences, all open programs on the PC were terminated.

The results obtained for the behavior models of MDT 1 to 4 are presented in Figure 15 as a bar chart. The mean times are represented by bars, and the maximum and minimum times are represented by error bars. The absolute range of times for MDT 1, 2 and 3 is minimal. In comparison, the absolute range of variation for MDT 4 is significantly larger. The results also demonstrate that the time required for compilation at MDT 4 represents a relatively minor component of the total time, accounting for approximately 12 % of the overall duration. In contrast, for MDT 1, 2 and 3, the time for compilation represents a significantly larger proportion of the total time, ranging from approximately 72 % to 76 %. However, in absolute terms, the compilation time for MDT 4 (26.78 s) is considerably higher than the time for MDT 1, 2 and 3 (0.76 s to 0.91 s). The reduction in computing time resulting from the abstraction from MDT 4 to 1, 2 or 3 is significant. In contrast, the difference in simulation times with and without compilation between MDT 1, 2 or 3 is minimal. While there is a slight increase from MDT 1 to 3, this is negligible at approximately 140 ms for the time with compilation. This trend is similarly observed for the time without compilation, with one exception for MDT 3. This exception is likely attributed to measurement errors in the very short measurement times.

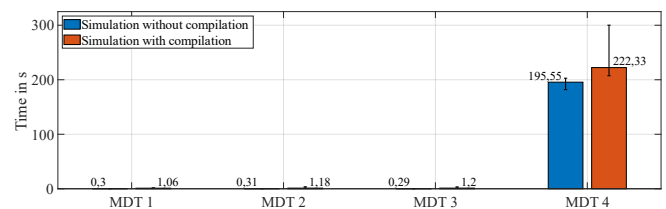

Figure 15: Times for the pure simulation and the simulation including compilation for the behavior model of MDT 1 to 4 of use case 1

A comparable pattern is observed in the memory requirements of the behavior models across different MDTs. This is illustrated in Table 3. The memory demand for the behavior models of MDT 1, 2 and 3 demonstrates a slight increase. Nevertheless, a notable rise is evident solely for MDT 4.

Table 3: Memory requirements for behavior models of MDTs 1 to 4 of use case 1

|  | MDT 1 | MDT 2 | MDT 3 | MDT 4 |
| --- | --- | --- | --- | --- |
| Storage size kb | 41 | 42 | 43 | 1.773 |

*B. Use case 2: Loading and unloading unit of a laser cutting machine*

The second industrial use case relates to a loading and unloading unit for laser cutting machines. Such a unit is used in laser cutting machines to achieve complete automation of the production process. The functions of the unit may be divided into two categories: the feeding of new sheets into the processing zone and the removal of finished cut components and scrap skeletons from the processing zone. When loading the laser cutting machine, large metal sheets are fed into the processing zone; when unloading, the cut-out sheet metal parts are taken out of the processing zone onto a transport carrier using flexible molds.

To optimize the efficiency of the development process for such machines, system manufacturers rely on the latest technological resources. One of the instruments used in this procedure is VC [15]. To create a virtual machine, the system manufacturer typically obtains the necessary behavior models of the components and systems from the component manufacturers. In this context, MDT 1, 2 and 3 are of particular relevance for the digital mapping of the control behavior of the components and systems [43]. The structure depicted in Figure 16 shows the loading and unloading unit of use case 2.

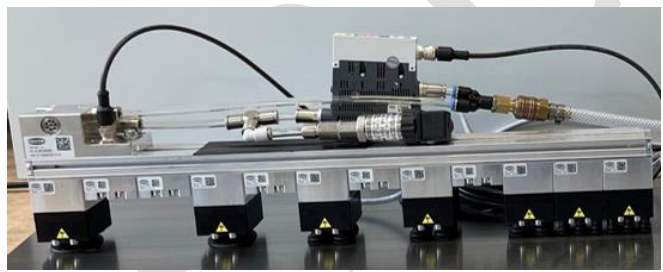

Figure 16: Loading and unloading unit with peripherals for measurement of use case 2

The unit itself is composed of a head module and a total of 12 add-on modules, which can have different configurations. For the purposes of this use case, 2 modules are equipped with 1 vacuum suction cup, 3 modules are equipped with 4 vacuum suction cups, and 2 modules are equipped with 9 vacuum suction cups. The remaining 5 module positions are equipped with so-called dummy modules, which are modules without vacuum suction cups. The device is controlled via IO-Link. This enables the individual control of each of the 32 vacuum suction cups. Pneumatic valves are utilized in the individual modules to control the vacuum suction cups. When activated, a vacuum suction cup is connected to the vacuum generator via a vacuum connection. Conversely, when the vacuum suction cup is deactivated, no vacuum flows through from the vacuum generator. Additionally, the modules indicate whether specific vacuum switching thresholds have been reached. Besides these vacuum switching thresholds, the head unit also reports other values, such as operating pressure or system vacuum.

Of all these values, only the switching thresholds H2 to H5 should be considered for the investigations in use case 2. The head unit is responsible for reporting the switching threshold designated as H2 centrally, while the remaining thresholds, H3, H4, and H5, are reported separately by each individual module. The standard values of the loading and unloading unit are used for the switching thresholds H2 to H5. These are listed in Table 4.

Table 4: Switching thresholds for the loading and unloading unit

| Switching threshold | H2 | H3 | H4 | H5 |
| --- | --- | --- | --- | --- |
| Value in -mbar,rel | 550 | 500 | 600 | 750 |

An ejector type SCPSi 10 [44] is used as the vacuum generator, controlled by a PLC. The PLC also controls the loading and unloading unit and records the vacuum values using an additional vacuum sensor located between the vacuum generator and the loading and unloading unit. The recorded values are obtained at a sampling rate of 1 ms. The documented measurement data is then compared with the output values of the behavior model of the system, which is created automatically using the AutoBMC method. A comparison of both curves is shown in Figure 17.

In the upper section, the vacuum curve is plotted over time, corresponding to the use cases 1. This demonstrates a high degree of correlation between the observed data and the predicted values, both for the range of increasing vacuum and for the range of constant vacuum and decreasing vacuum. The maximum deviation observed during the evacuation process was 47 mbar,rel, and 7 mbar,rel during the constant vacuum. The maximum deviation for the discarding process is 108 mbar,rel. The measurement curves for 30 measurement cycles are also presented in this diagram. As with the previously discussed use case, the scatter for this case is also minimal, which is why the maximum, minimum, and mean values in Figure 17 are almost congruent.

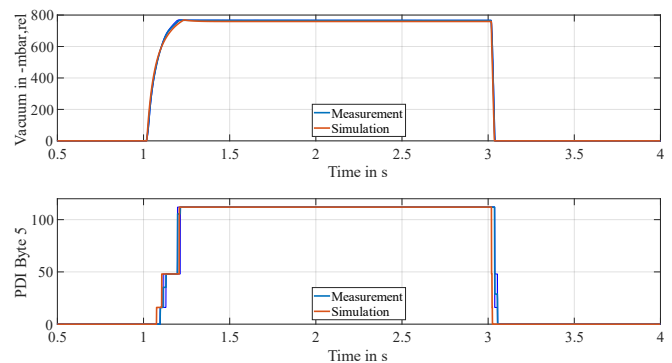

Figure 17: Comparison of measurement and simulation for use case 2a with vacuum curve (top) and output signal of PDI byte 5 (bottom)

A newly introduced feature for this use case is the display of an additional measured value. It describes the switching thresholds H3, H4, and H5, which have been measured in the individual modules that comprise the loading and unloading unit. The switching thresholds are expressed as a byte value, wherein the individual bits are assigned different functionalities. In this article, only bits 4 (H3), 5 (H4), and 6 (H5) are used. If a switching threshold is reached in the respective module, the bit is set to the value of 1. Conversely, if the threshold has not been reached, the bit is assigned the value of 0. To illustrate this, with a vacuum level of 650 mbar,rel in the module, bits 4 and 5 are set, resulting in a byte value of 48. The individual bytes are transmitted via IO-Link. It should be noted that the value shown explicitly describes the switching thresholds from the first module but is provided by the unit via byte number 5.

The correlation between measurement and simulation is particularly noteworthy for byte 5 (PDI byte 5). Note the significant discrepancy between the individual measurements and the vacuum curve. It is assumed that this phenomenon is due to the transmission mode of the signal (IO-Link), since the minimum cycle time for this protocol is 16.4 ms [45]The appearance of a switching threshold in relation to the cycle time can result in delays of at least 16.4 ms. There are significant discrepancies between the measured and simulated data and the vacuum curve, especially near second 3. This discrepancy can likely be attributed to the inherent limitations in the dynamics of IO-Link too, as the vacuum curve demonstrates a high degree of agreement between the measurement and simulation data. The same applies to the section of the falling vacuum as this is also very dynamic. Here too, the signal transmission causes relatively long delays.

The behavior model of the loading and unloading unit at MDT 4 exhibits a sufficiently accurate correspondence with the measurement data. However, the behavior models of MDT 1, 2 and 3 are of primary interest in this article. These are created entirely automatically with the presented method. The degree of agreement with the measurement is a crucial factor in the evaluation of the behavior models of MDT 1, 2 and 3. A comparison of the output signals produced by the behavior models of MDT 1, 2, 3 and 4 and the measurement is illustrated in Figure 18.

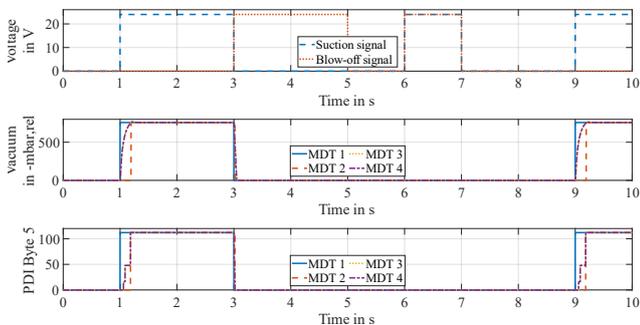

Figure 18: Behavior of the behavior models of MDT 1, 2 and 3 in comparison with MDT 4 for the signal H2 (bottom) and the system vacuum (middle) depending on the input signals suction and release (top) for use case 2

The suction and blow-off signal, as well as the vacuum curve, are equivalent to use case 1. The lower graph depicts the PDI byte 5. Similar to use case 1, the correlation between the output signals of MDT 1, 2 or 3 and those of MDT 4 is also highly significant for use case 2. The discrepancies between the individual MDTs are primarily attributable to the inherent limitations of the respective MDT.

However, it is also crucial for abstracted behavior models to be executed in a manner that conserves resources. To verify this claim, the behavior models of use case 2 are executed a total of 30 times at each MDT, like use case 1, and the times required for this are recorded. The results are presented in Figure 19.

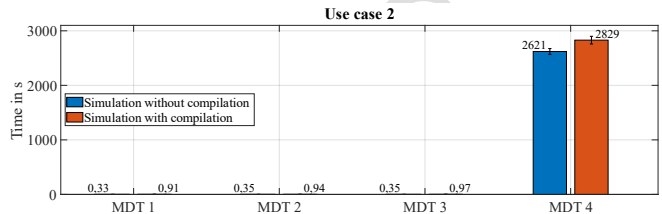

Figure 19: Times for the pure simulation and the simulation including compilation for the behavior models of the MDT 1 to 4 of use case 2

The results of use case 2 show that the behavior models of MDT 1, 2 and 3 require significantly less simulation time when both compilation and execution are enabled. MDT 4 behavior models can take a factor of about 2,000 longer to execute than MDT 1, 2 and 3 behavior models. The exact value depends on the exact times being compared.

As in the previous comparison, the execution time with and without compilation increases in proportion to the MDT of the behavior models. As with use case 1, the differences between MDT 1, 2 and 3 are marginal. A similar pattern emerges about the memory requirements of the behavior models of varying MDTs, as illustrated in Table 5. The memory requirement for the behavior models of MDT 1, 2 and 3 shows a slight increase with increasing MDT. However, there is a notable increase in this requirement for MDT 4.

Table 5: Memory requirements for behavior models of MDT 1 to 4 of use cases 2

|  | Use case 2 | | | |
| --- | --- | --- | --- | --- |
|  | MDT 1 | MDT 2 | MDT 3 | MDT 4 |
| Storage size kb | 52 | 52 | 54 | 1.682 |

*C. Discussion*

The automatically created behavior models of MDT 4 exhibit a high degree of correspondence with the real-life behavior in the 2 use cases. As the behavior models of MDT 1, 2 and 3 correspond closely with those of MDT 4, it can be concluded that they also align well with the real-life behavior. The primary limiting factor for the consistency between the various MDTs is the respective degree of simplification. However, the simplifications result in the exclusion of certain correlations from the model. Consequently, the abstracted behavior models are primarily suitable for applications where only a constrained model scope is necessary. This is frequently the case in larger overall systems or in time-critical applications. In such scenarios, abstracted behavior models can exhibit significant advantages, including reduced memory requirements and, most notably, accelerated execution times.

The creation of behavior models with a reduced MDT requires a significant amount of time, dependent on the dimensions of the behavior model, the inputs under consideration, and the value ranges. It is therefore recommended that the creation of abstracted behavior models be linked directly to tools for digital system design. It may be feasible to develop platforms that can be utilized directly by customers to fulfill their specific requirements. Subsequently, an optimized system is designed in accordance with the specified requirements using the behavior models of MDT 4. Once a suitable system has been selected, the corresponding behavior model can then be abstracted to MDT 1, 2 or 3. This process can be carried out fully automatically using suitable environments. After completion, the abstracted behavior models can be made available to the customer. In such a structure, the abstraction time would not be directly noticeable for the user, as abstracted behavior models are often not required directly when purchasing a system. The customer may then utilize the abstracted behavior model in their application (e.g. VC), where it can fully exploit the major advantages of the shorter simulation time.

The specific number of abstracted behavior models of MDT 1, 2 and 3 required is irrelevant for the duration of the abstraction. Most of the time required is dedicated to the capture of information regarding states, transitions, and other pertinent details, rather than the creation of the abstracted behavior models.

It should be noted, however, that the time required to manually create abstract behavior models is also a significant factor [46]. Manual creation of such models by experts is a considerably more time-consuming process, and it is also much more prone to errors[46].

## VI. CONCLUSION AND OUTLOOK

Behavior models for different use cases of the Digital Twin are typically required at different modeling depths depending on the available resources such as such as computing power, simulation time and memory requirements. This article presents a concept for the automated creation of behavior models at different modeling depths, with an evaluation based on 2 comprehensive industrial use cases in the domain of vacuum handling technology. The findings as followed:

- The behavior models created at modeling depth 4 exhibit a high degree of correlation with the observed behavior of the investigated systems.

- The behavior models created for modeling depths 1, 2 and 3 exhibit a high degree of correlation with modeling depth 4 and, by extension, with the measured curves. However, the degree of correlation is limited by the specific parameters and constraints associated with each modeling depth.

- A reduction in the modeling depth of a behavior model results in an increase of the execution speed. While the discrepancy between modeling depths 1, 2 and 3 remains relatively minor, it increases significantly for behavior models with modeling depth 4. This phenomenon also manifests in the memory requirements of individual behavior models.

As the modeling depth decreases, the memory requirement also decreases, with the changes between modeling depths 1, 2, and 3 being significantly smaller compared to modeling depth 4.

In addition to applying the presented concept to other domains, a further reduction of the duration for detecting the relevant states and transitions is useful, especially for larger behavior models. Furthermore, a detailed comparison of the presented concept with approaches from literature, using the example of different use cases, can clarify the advantages and disadvantages of the presented concept even better.

ACKNOWLEDGMENT

This work was supported by the Ministry of Science, Research and the Arts of the State of Baden-Wurttemberg within the sustainability support of the projects of the Excellence Initiative II.

REFERENCES

[1] E. Westkämper and C. Löffler, *Strategien der Produktion: Technologien, Konzepte und Wege in die Praxis*: Springer Berlin Heidelberg, 2016.
[2] E. Negri, L. Fumagalli, and M. Macchi, "A Review of the Roles of Digital Twin in CPS-based Production Systems," *Procedia Manufacturing*, vol. 11, pp. 939–948, 2017, doi: 10.1016/j.promfg.2017.07.198.
[3] M. Sjarov et al., "The Digital Twin Concept in Industry – A Review and Systematization," in *2020 25th IEEE International Conference on Emerging Technologies and Factory Automation (ETFA)*, Vienna, Austria, Sep. 2020 - Sep. 2020, pp. 1789–1796.
[4] Z. Sun, R. Zhang, and X. Zhu, "The progress and trend of digital twin research over the last 20 years: A bibliometrics-based visualization analysis," *Journal of Manufacturing Systems*, vol. 74, pp. 1–15, 2024, doi: 10.1016/j.jmsy.2024.02.016.
[5] W. Kritzinger, M. Karner, G. Traar, J. Henjes, and W. Sihn, "Digital Twin in manufacturing: A categorical literature review and classification," *IFAC-PapersOnLine*, vol. 51, no. 11, pp. 1016–1022, 2018, doi: 10.1016/j.ifacol.2018.08.474.
[6] A. Fuller, Z. Fan, C. Day, and C. Barlow, "Digital Twin: Enabling Technologies, Challenges and Open Research," *IEEE Access*, vol. 8, pp. 108952–108971, 2020, doi: 10.1109/ACCESS.2020.2998358.
[7] B. Ashtari Talkhestani et al., "An architecture of an Intelligent Digital Twin in a Cyber-Physical Production System," *at - Automatisierungstechnik*, vol. 67, no. 9, pp. 762–782, 2019.
[8] L. Wright and S. Davidson, "How to tell the difference between a model and a digital twin," *Adv. Model. and Simul. in Eng. Sci.*, vol. 7, no. 1, 2020, doi: 10.1186/s40323-020-00147-4.
[9] B. Ashtari Talkhestani et al., "An architecture of an Intelligent Digital Twin in a Cyber-Physical Production System," *at - Automatisierungstechnik*, vol. 67, no. 9, pp. 762–782, 2019.
[10] P. Barabinot, R. Scanff, P. Ladevèze, D. Néron, and B. Cauville, "Industrial Digital Twins based on the non-linear LATIN-PGD," *Adv. Model. and Simul. in Eng. Sci.*, vol. 8, no. 1, 2021, doi: 10.1186/s40323-021-00207-3.
[11] R. Rosen et al., "Die Rolle der Simulation im Kontext des Digitalen Zwillings," *atp*, vol. 63, no. 04, pp. 82–89, 2021, doi: 10.17560/atp.v63i04.2534.
[12] S. Boschert and R. Rosen, "Digital Twin—The Simulation Aspect," in *Mechatronic Futures: Challenges and Solutions for Mechatronic Systems and their Designers*, P. Hehenberger and D. Bradley, Eds., Cham, s.l.: Springer International Publishing, 2016, pp. 59–74.
[13] M. Lochbichler, *Systematische Wahl einer Modellierungstiefe im Entwurfsprozess mechatronischer Systeme*. Paderborn: Universität Paderborn Heinz Nixdorf Institut, 2020.
[14] K. Kleeberger, J. Schnitzler, M. U. Khalid, R. Bormann, W. Kraus, and M. F. Huber, "Precise Object Placement with Pose Distance Estimations for Different Objects and Grippers," 2021.
[15] N. Striffler and T. Voigt, "Concepts and trends of virtual commissioning – A comprehensive review," *Journal of Manufacturing Systems*, vol. 71, pp. 664–680, 2023, doi: 10.1016/j.jmsy.2023.10.013.

[16] D. Dittler, V. Stegmaier, N. Jazdi, and M. Weyrich, "Illustrating the benefits of efficient creation and adaption of behavior models in intelligent Digital Twins over the machine life cycle," Jun. 2024. [Online]. Available: http://arxiv.org/pdf/2406.08323v1

[17] V. Stegmaier, D. Dittler, N. Jazdi, and M. Weyrich, "A structure of modelling depths in behavior models for Digital Twins," submitted, 2022.

[18] H. Kuolt, J. Gauß, W. Schaaf, and A. Winter, "Optimization of pneumatic vacuum generators: heading for energy-efficient handling processes," in *10th International Fluid Power Conference*, Dresden, 2016, pp. 267–280.

[19] D. Dittler, P. Lierhammer, D. Braun, T. Müller, N. Jazdi, and M. Weyrich, "A Novel Model Adaption Approach for intelligent Digital Twins of Modular Production Systems," in *2023 IEEE 28th International Conference on Emerging Technologies and Factory Automation (ETFA)*, Sinaia, Romania, 9122023, pp. 1–8.

[20] V. Stegmaier, W. Schaaf, N. Jazdi, and M. Weyrich, "Efficient Creation of Behavior Models for Digital Twins Exemplified for Vacuum Gripping Systems," in *2022 IEEE 27th International Conference on Emerging Technologies and Factory Automation (ETFA)*, Stuttgart, Germany, 2022, pp. 1–8.

[21] D. Straub and W. Schaaf, "Experimental and Theoretical Investigation of Lightweight Pumps and Fluid Reservoirs for Electrically Driven Vacuum Systems in Automated Handling Processes," in *11th International Fluid Power Conference*, Aachen, 2018, pp. 434–445.

[22] D. Straub, *Methode zur technischen Auslegung von Vakuumgreifsystemen mit einer Mindesthaltedauer auf Basis fluidischer Untersuchungen*. Stuttgart: Fraunhofer Verlag, 2021.

[23] K. S. Mohamed, *Machine Learning for Model Order Reduction*. Cham: Springer International Publishing, 2018.

[24] P. Benner, T. Breiten, H. Faßbender, M. Hinze, T. Stykel, and R. Zimmermann, *Model Reduction of Complex Dynamical Systems*. Cham: Springer International Publishing, 2021.

[25] I. The MathWorks, *Reduced Order Modeling*. [Online]. Available: https://de.mathworks.com/discovery/reduced-order-modeling.html (accessed: Mar. 23 2024).

[26] L. d. P. P. Mapa, F. d. A. das Neves, and G. P. Guimarães, "Dynamic Substructuring by the Craig–Bampton Method Applied to Frames," *J. Vib. Eng. Technol.*, vol. 9, no. 2, pp. 257–266, 2021, doi: 10.1007/s42417-020-00223-4.

[27] K. Mohaghegh, "Linear and nonlinear model order reduction for numerical simulation of electric circuits," Dissertation, Lehrstuhl für Angewandte Mathematik, Bergische Universität Wuppertal, Wuppertal, 2010.

[28] J. Bokor and G. Balas, "LINEAR PARAMETER VARYING SYSTEMS: A GEOMETRIC THEORY AND APPLICATIONS," *IFAC Proceedings Volumes*, vol. 38, no. 1, pp. 12–22, 2005, doi: 10.3182/20050703-6-CZ-1902.00003.

[29] M. Heinkenschloss, T. Reis, and A. C. Antoulas, "Balanced truncation model reduction for systems with inhomogeneous initial conditions," *Automatica*, vol. 47, no. 3, pp. 559–564, 2011, doi: 10.1016/j.automatica.2010.12.002.

[30] T. Simpson, N. Dervilis, and E. Chatzi, "Machine Learning Approach to Model Order Reduction of Nonlinear Systems via Autoencoder and LSTM Networks," *J. Eng. Mech.*, vol. 147, no. 10, p. 9, 2021, doi: 10.1061/(ASCE)EM.1943-7889.0001971.

[31] H. F. S. Lui and W. R. Wolf, "Construction of Reduced Order Models for Fluid Flows Using Deep Feedforward Neural Networks," *J. Fluid Mech.*, vol. 872, pp. 963–994, 2019, doi: 10.1017/jfm.2019.358.

[32] C. J. G. Rojas, A. Dengel, and M. D. Ribeiro, "Reduced-order Model for Fluid Flows via Neural Ordinary Differential Equations," Feb. 2021. [Online]. Available: http://arxiv.org/pdf/2102.02248v2

[33] MoRePaS, *RBmatlab*. [Online]. Available: https://www.morepas.org/software/rbmatlab

[34] CADFEM Germany GmbH, *MODEL REDUCTION INSIDE ANSYS – MODEL REDUCTION FOR FASTER CALCULATIONS*. [Online]. Available: https://www.cadfem.net/en/our-solutions/cadfem-ansys-extensions/model-reduction-inside-ansys.html (accessed: Mar. 23 2024).

[35] *pyMOR - Model Order Reduction with Python*. [Online]. Available: https://github.com/pymor/pymor?tab=readme-ov-file (accessed: Mar. 23 2024).

[36] P. Benner and S. W. R. Werner, *MORLAB - Model Order Reduction LABoratory*. [Online]. Available: http://www.mpi-magdeburg.mpg.de/projects/morlab (2019). doi:10.5281/zenodo.3332716

[37] V. Stegmaier, W. Schaaf, N. Jazdi, and M. Weyrich, "Simulation Model for Digital Twins of Pneumatic Vacuum Ejectors," *Chem Eng & Technol*, vol. 46, no. 1, pp. 71–79, 2023, doi: 10.1002/ceat.202200358.

[38] J. Schmalz GmbH, *SBP 20 S03 SDA: Art-Nr.: 10.02.01.00567*. Grundejektor für universellen Einsatz. [Online]. Available: https://www.schmalz.com/de-de/vakuumtechnik-fuer-die-automation/vakuum-komponenten/vakuum-erzeuger/grundejektoren/grundejektoren-sbp-307660/10.02.01.00567/ (accessed: Nov. 27 2023).

[39] T. Eberhardt, V. Stegmaier, W. Schaaf, and A. Verl, "A TRAJECTORY-SPECIFIC APPROACH FOR CALCULATING THE REQUIRED HOLDING FORCE FOR SURFACE GRIPPERS," in *14th International Fluid Power Conference*, Dresden, Germany, 2024, pp. 623–635.

[40] V. Stegmaier, G. Ghasemi, N. Jazdi, and M. Weyrich, "An approach enabling Accuracy-as-a-Service for resistance-based sensors using intelligent Digital Twins," *Procedia CIRP*, vol. 107, pp. 833–838, 2022, doi: 10.1016/j.procir.2022.05.071.

[41] D. Straub and K. Huber, "Potentials of vacuum gripping systems in human-robot-collaboration applications," in *ISR 2018: 50th International Symposium on Robotics*, 2018, pp. 155–158.

[42] F. Gabriel, P. Bobka, and K. Dröder, "Model-Based Design of Energy-Efficient Vacuum-Based Handling Processes," *Procedia CIRP*, vol. 93, pp. 538–543, 2020, doi: 10.1016/j.procir.2020.03.006.

[43] V. Stegmaier, D. Dittler, N. Jazdi, and M. Weyrich, "A Structure of Modelling Depths in Behavior Models for Digital Twins," in *2022 IEEE 27th International Conference on Emerging Technologies and Factory Automation (ETFA)*, Stuttgart, Germany, 962022, pp. 1–8.

[44] J. Schmalz GmbH, *SCPSi 10 G02 NC M12-5: Art-Nr.: 10.02.02.04123*. Kompaktejektor m. integrierter Luft- sparregelung in schmaler Bauform und IO-Link Funktionalität. [Online]. Available: https://www.schmalz.com/de-de/vakuumtechnik-fuer-die-automation/vakuum-komponenten/vakuum-erzeuger/kompaktejektoren/kompaktejektoren-scps-scpsi-307842/10.02.02.04123/

[45] Pascal Gaggero, *Real-time speed of IO-Link wired*. [Online]. Available: https://www.innovating-automation.blog/real-time-speed-of-io-link/ (accessed: Mar. 15 2024).

[46] Valentin Stegmaier, Tobias Eberhardt, Nasser Jazdi, and Michael Weyrich, "Quantitative Evaluation of Automated Behavior Model Creation for Applications of Industrial Automation," 2024.